\documentclass[superscriptaddress,twocolumn]{revtex4-2}
\usepackage{amsmath}
\usepackage{graphicx}
\usepackage{dcolumn}
\usepackage{bm}
\usepackage{color}
\usepackage{soul}
\usepackage{multirow}
\usepackage{ulem}
\usepackage[marginal]{footmisc}
\usepackage[colorlinks,linkcolor=blue,anchorcolor=blue,citecolor=blue,urlcolor=blue]{hyperref}
\usepackage{soul}
\usepackage{ulem}
\usepackage{xcolor}
\DeclareRobustCommand{\erase}{\bgroup\markoverwith{\textcolor{red}{\rule[.5ex]{2pt}{0.4pt}}}\ULon}
\newcommand{\be}{\begin{equation}}
\newcommand{\ee}{\end{equation}}
\newcommand{\bea}{\begin{eqnarray}}
\newcommand{\eea}{\end{eqnarray}}
\renewcommand{\vec}[1]{\mbox{\boldmath $#1$}}

\begin{document}
\title{Proton-neutron pairing correlations in N=Z nuclei with Deformed Skyrme+pnBCS model}

\author{Xin Lian}
\affiliation{College of Physics, Sichuan University, Chendu 610065, China}

\author{C. L. Bai}
\affiliation{College of Physics, Sichuan University, Chendu 610065, China}

\author{H. Sagawa}
\affiliation{Center for Mathematics and Physics, University of Aizu, Aizu-Wakamatsu, Fukushima 965-8560, Japan}
\affiliation{RIKEN, Nishina Center, Wako 351-0198, Japan}

\author{H. Q. Zhang}
\affiliation{China Institute of Atomic Energy, Beijing 102413, China}

\begin{abstract}
We investigate the effects of neutron-neutron (nn), proton-proton (pp) and proton-neutron (pn) pairing correlations  on the ground-states of $N = Z$
even-even $pf$-shell nuclei 
by using an axially symmetric deformed Hartree-Fock (HF)+pnBardeen-Cooper-Schrieffer (BCS) model.   
We adopt a Skyrme energy density functional (EDF) SGII,  together with contact volume- and surface-type pairing interactions, whose strengths are adjusted to 
reproduce empirical pairing gaps of each nucleus.  
It is shown that the strength of the IS pairing is correlated to the nuclear deformation: for oblate deformation with $-0.3<\beta_2<0.0$,  a stronger IS pairing 
is required to reproduce the empirical pairing gap, 
while for prolate deformation a weaker one is enough. 
Among the eight $N=Z$ nuclei, we found that $^{64}$Ge, $^{68}$Se and $^{72}$Kr show the dominance of isovector (IV) spin-singlet superfluidity, while lighter 5 nuclei $^{44}$Ti, $^{48}$Cr, $^{52}$Fe, $^{56}$Ni and $^{60}$Zn  exhibit the coexistence of IV spin-singlet and isoscalar (IS) spin-triplet  superfluidities.
We found also that the IS abnormal density always exhibits the oblate deformation, regardless of whether the normal density is prolate, spherical  or oblate.
\end{abstract}

\maketitle

\section{introduction} 
The proton-neutron (pn) pairing becomes active in N$\sim$Z nuclei where valence protons and neutrons occupy the same orbitals or 
the spin-orbit partners, having the maximum overlap between their wave functions \cite{Frauendor,Sagawa16}.
The pn pair can be further divided into isoscalar spin-triplet (IS) and isovector spin-singlet (IV) pairing components according to the total isospin of two nucleons. 
The pn pairing, particularly IS pairing correlations have significant impacts on the properties of $N\sim Z$ nuclei, such as the origin of Wigner energy \cite{Satula,Bentley2013,Bentley2014},
the existence of quartet cluster\cite{Sandulescu2015,Sandulescu2012(2),Sandulescu2012(1),Sandulescu2014}, 
the abnormal bifurcation of the double binding 
energy differences between the odd-odd and even-even nuclei\cite{ZhaoPW}, 
low-lying spectrum of excited states\cite{Qi11}, 
the strong low-energy Gamow-Teller(GT) transition \cite{Bai2013,Fujita,Bai2014},
and the deuteron transfer reaction \cite{Isacker,Yoshida}.
In nuclei with neutron excess, the IS pairing 
has also significant effects on $\beta$ decay \cite{Engel99,Niksic} and double $\beta$ decay \cite{Engel2008,Sorensen} induced by low-energy GT strength below the $\beta$ decay $Q$ value.
Experimentally, the data in $^{92}$Pd suggest a strong IS pairing correlation different from the normal IV superfluidity~\cite{Cederwall}.
In the near future, high-intensity radioactive beams will provide new possibilities for studying the structural properties of unstable nuclei along the $N=Z$ 
line enhancing the possibility of finding new evidences for pn  pairing \cite{Frauendor}.

In order to explore the effects of pn pairing on the ground state properties, the Hartree-Fock-Bogoliubov (HFB) and Bardeen-Cooper-Schrieffer(BCS) including
the pn pairing (pnHFB or pnBCS) were introduced in Refs. \cite{Goodman1998,Goodman1972,Goodman1999,Bulthuis,Bertsch2010,Bertsch2011,Gezerlis2025,Simkovic,Cheoun2015,Cheoun2018(1),Cheoun2018(2)}.
Later, the self-consistent mean-field calculations in the full valence space were done, in which the paring interactions were taken in the forms of 
volume force \cite{Bulthuis,Bertsch2010,Bertsch2011,Gezerlis2025}, schematic force \cite{Simkovic}, and the  realistic CD Bonn potential \cite{Cheoun2015,Cheoun2018(1),Cheoun2018(2)}.
In the above literatures, the obvious coexistence of IS and IV pn pairing correlation in $N=Z$ nuclei can be obtained only when the  realistic pairing interactions 
were adopted \cite{Goodman1999,Cheoun2015,Cheoun2018(1),Cheoun2018(2)}, which enable to reproduce empirical nn, pp, and pn  paring gaps, simultaneously.
Moreover, IS pairing has been suggested to exist close to the N=Z line near A$\sim$130 in Refs.\cite{Bertsch2011,Gezerlis2025}, which shows the deformation plays an important role.
And in $sd$-shell nuclei when varying the IS and IV pairing strengths with a  fixed ratio of 1.5, the IS pairing dominates for large prolate or oblate deformations with $|\beta_2|>0.3$ in $^{24}$Mg and $^{28}$Si, and only for large prolate 
deformation with $\beta_2>0.3$ in $^{32}$S \cite{Cheoun2018(1)}. For smaller deformations, it is still ambiguous whether
a strong IS correlation exists.

In this article,  we study the co-existence problem of IS and IV superfluidities in even-even N=Z
 $pf-$shell nuclei with the self-consistent HF+pnBCS calculations including the freedom of axial-symmetric quadrupole deformation using Skyrme EDFs. 
As the pairing interactions, the volume and surface pairing interactions are employed for $pp, nn$ and $pn$ channels, and  the IS and IV pairing strengths are varied independently to reproduce the empirical pairing gaps.
The coexistence of IS and IV pairing condensations will be examined under the presence of the quadrupole deformation. 
This paper is organized as follows. In Section \ref{section:II}, we present detailed formalism of the deformed pnBCS model. Numerical results and related discussions are presented in Section \ref{section:III}. Finally, a summary is presented in Section \ref{section:IV}.

\section{formalism and details}\label{section:II}
\subsection{BCS model including pn pairing}
The  IS and IV pairing correlations on the ground state are accounted by including all pp, nn and pn pairing interactions in a BCS model so called "pnBCS" model  \cite{Goswami}.
In this model, the Hamiltonian of the system is composed from the one-body part $H_0$ and two-body part $H_{int}$, 
\begin{equation}
\begin{split}
H&=H_0+H_{int}\\
H_0&=\sum_{\alpha q}(\varepsilon_{\alpha q} -\lambda _{q})C_{\alpha q}^{\dag}C_{\alpha q} \\
H_{int}&=\frac{1}{2}\sum_{\substack{\alpha,q_1,q_2\\\beta,q_3,q_4}}\upsilon_{(\alpha q_1)(\bar{\alpha}q_2)(\beta q_3)(\bar{\beta}q_4)}C^{\dag}_{\alpha q_1}C^{\dag}_{\bar{\alpha}q_2}C_{\bar{\beta}q_4}C_{\beta q_3},\label{vint}
\end{split}
\end{equation}
where $\alpha$ and $\beta$ denote the single-particle states with quantum number $\Omega^\pi$ of the $z$ component of angular momentum and parity in the axially symmetric deformed potential, and $q=(n, p)$ denotes neutron or proton.  $\bar{\alpha}$ and $\bar{\beta}$ denote the time-reversed state of $\alpha$ and $\beta$, respectively.
$\varepsilon_{\alpha q}$ represents the single-particle energy, and $C_{\alpha q}^{\dag}$ ($C_{\alpha q}$) is the corresponding creation (annihilation) operator, while  $\lambda _{q}$ is the chemical potential. $\upsilon_{(\alpha q_1)(\bar{\alpha}q_2)(\beta q_3)(\bar{\beta}q_4)}$ is the two-body matrix element for the pairing interaction.
Assuming the time-reversal symmetry, the quasiparticle creation and annihilation operators are constructed by the following  transformation:
\begin{eqnarray}
\begin{split}
\left(\begin{array}{c} a_{\alpha1}^{\dag}\\a_{\alpha 2}^{\dag}\\a_{\alpha \overline{1}}\\a_{\alpha \overline{2}}\end{array}\right)= 
\left(\begin{array}{cccc}
u_{\alpha 1p} & u_{\alpha 1n} & v_{\alpha 1p} & v_{\alpha 1n}\\
u_{\alpha 2p} & u_{\alpha 2n} & v_{\alpha 2p} & v_{\alpha 2n}\\
-v_{\alpha 1p} & -v_{\alpha 1n} & u_{\alpha 1p} & u_{\alpha 1n}\\
-v_{\alpha 2p} & -v_{\alpha 2n} & u_{\alpha 2p} & u_{\alpha 2n}
\end{array}\right)
\left(\begin{array}{c} C_{\alpha p}^{\dag}\\C_{\alpha n}^{\dag}\\ C_{\alpha \overline{p}}\\C_{\alpha \overline{n}}\end{array}\right),
\end{split}
\end{eqnarray}
where $i=1$ or $2$ labels the isospin of the quasiparticles. 
The superposition factors $u$ and $v$ can be obtained solving the pnBCS equation,
\begin{eqnarray}
&&\left(\begin{array}{cccc}
\varepsilon_{\alpha p}-\lambda_{p}&0&\triangle_{\alpha p \overline{p}}&\triangle_{\alpha p \overline{n}} \\
0&\varepsilon_{\alpha n}-\lambda_{n}&\triangle_{\alpha n \overline{p}}&\triangle_{\alpha n \overline{n}}\\
\triangle_{\alpha p \overline{p}}&\triangle_{\alpha p \overline{n}}&\lambda_{p}-\varepsilon_{\alpha p}&0\\
\triangle_{\alpha n \overline{p}}&\triangle_{\alpha n \overline{n}}&0&\lambda_{n}-\varepsilon_{\alpha n}\end{array}\right)
\left(\begin{array}{c} u_{\alpha ip}\\
u_{\alpha in}\\v_{\alpha i p}\\v_{\alpha i n}
\end{array}\right)\nonumber\\
&&=E_{\alpha i}\left(\begin{array}{c} u_{\alpha ip}\\
u_{\alpha in}\\v_{\alpha i p}\\v_{\alpha i n}
\end{array}\right).\label{bcs}
\end{eqnarray}
 Notice that the quasi-particle state has the good isospin, but mixes the proton and neutron states.

The pairing interactions for different  channels are given by, 
\begin{eqnarray}
V_j(\vec{r_1},\vec{r_2})&&=V_0^j(1-x_j\frac{\rho(\vec{r})}{\rho_0})\delta(\vec{r_1}-\vec{r_2})P_{j},\label{interaction}
\end{eqnarray}
where $j$ denotes pp, nn, IS and IV pn pairing with $P_{j}$ being the corresponding projection operators. $V^{j}_0$ are the strength parameters of the corresponding pairing channels, $x_j$ takes 0 or 1 corresponding to volume or surface parings, and $\rho_0=0.16$ fm$^{-3}$ is the saturation density. It should be noted that we allow different parameters for different paireding channels in our calculations. 

Under the axially symmetric deformation, the single-particle wave function $\Phi_{\alpha,q}$ and its time reversal are written as, 
\begin{equation}
\begin{split}
&\Phi_{\alpha,q}=\sum\limits_\sigma \phi_{\alpha\sigma q}( r_\perp,z)e^{i(\Omega-\sigma)\varphi}\chi_\sigma\chi_q,\\
&\Phi_{\bar{\alpha},q}=\sum\limits_\sigma(-1)^{\frac{1}{2}+\sigma}\phi_{\alpha-\sigma q}( r_\perp,z)e^{-i(\Omega+\sigma)\varphi}\chi_\sigma\chi_q,
\end{split}
\end{equation}
where $\phi_{\alpha\sigma q}( r_\perp,z)$ is the radial wave function in the cylindrical coordinate space $(r_\perp, z)$,  and $\sigma = \pm\frac{1}{2}$ label the up and down components of the spin. 
The pairing matrix elements in Eq.(\ref{bcs}) can then be calculated in the cylindrical coordinate space directly,
\begin{equation}
\begin{split}
\Delta_{\alpha p\overline{p}}=&-\sum \limits_{\beta>0,\sigma_1,\sigma_2}(u_{\beta 1p}v_{\beta 1p}^*+u_{\beta 2p}v_{\beta 2p}^*)\times\\ 
&\int\int |\phi_{\alpha\sigma_1 p} (r_\perp,z)|^2V_{pp} (r_\perp,z)|\phi_{\beta\sigma_2 p} (r_\perp,z)|^22\pi  r_\perp dr_\perp dz \\
\Delta_{\alpha n\overline{n}}=&-\sum \limits_{\beta>0,\sigma_1,\sigma_2}(u_{\beta 1n}v_{\beta 1n}^*+u_{\beta 2n}v_{\beta 2n}^*)\times\\ 
&\int\int |\phi_{\alpha\sigma_1 n} (r_\perp,z)|^2V_{nn} (r_\perp,z)|\phi_{\beta \sigma_2 n} (r_\perp,z)|^22\pi  r_\perp dr_\perp dz  \\
\Delta_{\alpha p\overline{n}}=&-\sum \limits_{\beta>0,\sigma_1,\sigma_2}Re(u_{\beta 1p}v_{\beta 1n}^*+u_{\beta 2p}v_{\beta 2n}^*)\\ 
&\times\int\int \phi_{\alpha\sigma_1 p} (r_\perp,z)\phi_{\alpha\sigma_1 n} (r_\perp,z)V_{IV} (r_\perp,z)\\
&\times \phi_{\beta \sigma_2 p} (r_\perp,z)\phi_{\beta \sigma_2 n} (r_\perp,z)2\pi  r_\perp dr_\perp dz\\
&-i\sum \limits_{\beta>0,\sigma_1,\sigma_2}Im(u_{\beta 1p}v_{\beta 1n}^*+u_{\beta 2p}v_{\beta 2n}^*)\\ 
&\times(-1)^{1-\sigma_1-\sigma_2}\int\int \phi_{\alpha\sigma_1 p} (r_\perp,z)\phi_{\alpha\sigma_1 n} (r_\perp,z)V_{IS} (r_\perp,z)\\
&\times \phi_{\beta \sigma_2 p} (r_\perp,z)\phi_{\beta \sigma_2 n} (r_\perp,z)2\pi  r_\perp dr_\perp dz,
\end{split}\label{dpn}
\end{equation}

In the current pnBCS scheme, only $ \overline{p}n$ and $p\overline{n}$ pairing correlations are included. 
The effects of $(pn)$ and $(\overline{p} \overline{n})$ components are effectively incorporated by multiplying a factor of $2$ on the IS component as was suggested in Ref. \cite{Cheoun2018(1)}.

\subsection{Pairing Gaps}
The empirical pp, nn and pn pairing gaps  are evaluated by the 
formulas \cite{Wu16,Yang22}, 
\begin{equation}
\begin{split}
\Delta^{emp}_{p}=&\frac{1}{8}[M(Z+2,N)-4M(Z+1,N)+6M(Z,N) \\ 
&-4M(Z-1,N)+M(Z-2,N)]\\
\Delta^{emp}_{n}=&\frac{1}{8}[M(Z,N+2)-4M(Z,N+1)+6M(Z,N)\\
&-4M(Z,N-1)+M(Z,N-2)]\\
\Delta^{emp}_{pn}=&\frac{1}{4}\lbrace2[M(Z,N+1)+M(Z,N-1)M(Z+1,N)\\&+M(Z-1,N)]-[M(Z+1,N+1)
\\&+M(Z-1,N+1)-M(Z+1,N-1)
\\&+M(Z-1,N-1)]-4M(Z,N) \rbrace
\end{split}.\label{egap}
\end{equation}

The empirical gaps of $N=Z$ even-even nuclei from $^{44}$Ti to $^{72}$Kr are listed in Table \ref{tab1}.
\begin{table}[h]
\caption{\label{tab1}Empirical pairing gaps of N = Z nuclei from $^{44}$Ti to $^{72}$Kr }.
\begin{ruledtabular}
\begin{tabular}{cccc}
\textrm{Nucleus}&
\textrm{$\Delta^{emp}_{p}$}&
\textrm{$\Delta^{emp}_{n}$}&
\textrm{$\Delta^{emp}_{pn}$}\\
\colrule
$^{44}$Ti&2.631&2.653&2.068\\
$^{48}$Cr&2.128&2.138&1.442\\
$^{52}$Fe&1.991&2.007&1.122\\
$^{56}$Ni&2.080&2.159&1.115\\
$^{60}$Zn&1.679&1.782&1.044\\
$^{64}$Ge&1.807&2.141&1.435\\
$^{68}$Se&1.909&2.184&1.522\\
$^{72}$Kr&2.001&1.985&1.353\\
\end{tabular}
\end{ruledtabular}
\end{table}

 Theoretically, the mean pairing gaps are evaluated by the following formulas,
\begin{flalign}
\begin{split}
&\overline{\Delta}_{pp}=\sum_{\alpha}\frac{\Delta_{ \alpha p\overline{p}}\langle S^{\dag}_{\alpha p\overline{p}}S_{\alpha p\overline{p}} \rangle}{\sum_{\beta}\langle S^{\dag}_{\beta p\overline{p}}S_{\beta p\overline{p}} \rangle}\\
&\overline{\Delta}_{nn}=\sum_{\alpha}\frac{\Delta_{ \alpha n\overline{n}}\langle S^{\dag}_{\alpha n\overline{n}}S_{\alpha n\overline{n}} \rangle}{\sum_{\beta}\langle S^{\dag}_{\beta n\overline{n}}S_{\beta n\overline{n}} \rangle} \\
&\overline{\Delta}_{pn}^{IV}=\sum_{\alpha}\frac{Re( \Delta_{ \alpha p\overline{n}}) \langle S^{\dag}_{\alpha p\overline{n}}S_{\alpha p\overline{n}}\rangle  }{\sum_{\beta} \langle S^{\dag}_{\beta p\overline{n}}S_{\beta p\overline{n}}\rangle }\\
&\overline{\Delta}_{pn}^{IS}=\sum_{\alpha}\frac{Im( \Delta_{ \alpha p\overline{n}}) \langle S^{\dag}_{\alpha p\overline{n}}S_{\alpha p\overline{n}}\rangle  }{\sum_{\beta} \langle S^{\dag}_{\beta p\overline{n}}S_{\beta p\overline{n}}\rangle }\\
&\overline{\Delta}_{pn}=\sum_{\alpha}\frac{\vert \Delta_{ \alpha p\overline{n}}\vert \langle S^{\dag}_{\alpha p\overline{n}}S_{\alpha p\overline{n}}\rangle  }{\sum_{\beta} \langle S^{\dag}_{\beta p\overline{n}}S_{\beta p\overline{n}}\rangle },
\end{split}\label{mgap}
\end{flalign}
where $S^{\dag}_{\alpha q_1 \overline{q_2}}=C_{\alpha q_1}^{\dag}C_{\alpha \overline{q_2}}^{\dag}$ is the pair creation operator.

Before the pnBCS calculation,  axially symmetric deformed HFB \cite{Stoitsov} calculations with  only pp and nn pairing interactions  are done to obtain the  single-particle states for the pnBCS calculation.  
In the HFB calculations, the model space for all nuclei is exploited up to $N = 6\hbar\omega$ in a deformed harmonic oscillator basis.
The deformation parameter $\beta_2$ of the harmonic oscillator basis and its corresponding mass quadratic moment are varied in the range $-0.5<\beta_2<0.5$. The strength parameters for the pp and nn pairing interactions are determined by fitting to the empirical pairing gaps. After the HFB calculations are done, the single-particle states in the canonical basis are extracted and used as the input for the pnBCS calculations, which are performed 
at each deformation point taking into account the full four pairing interactions.   Notice that the canonical basis is a substitute of deformed Woods-Saxon wave function as the initial wave functions  to make good convergence for pnBCS calculations.
The calculation is started with initial appropriate $pp, nn,$ and pn  gaps, and the iteration is finished when the self-consistent solution is obtained.

\section{results}\label{section:III}

In the present research, the Skyrme Energy Density Functional(EDF) SGII \cite{SGII} 
is employed.  We employed the  volume-type  pairing interaction for  all  four pairing channels in Section IIIA. Then in Section IIIB, we take the surface-type pairing interaction for the pp and nn channels,  and the volume-type one for the pn channel.
\subsection{Results with the volume-type interaction}
\begin{figure}[t]
  \includegraphics[scale=0.7, clip=true]{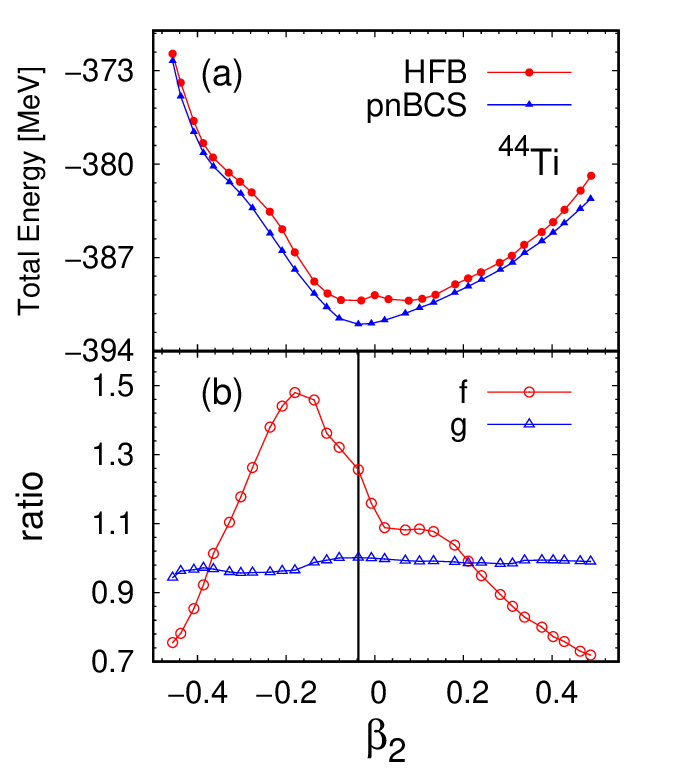}
  \caption{\label{oldratio} (a): The total energy of ground state of  $^{44}$Ti   as a function of the quadrupole deformation parameter $\beta_2$  
  calculated by the HFB and pnBCS models. The HFB includes only the IV pp and nn pairing interactions, while pnBCS takes into account  also IS and IV pn pairing ones.  Assuming $V^{pp}_0=V^{nn}_0=V^{IV}_0$  and $x_j=0$ in Eq. (\ref{interaction}), 
  the IS and IV pairing strengths at each deformation are adjusted to reproduce the empirical paring gaps in Table I.  (b): The paring strength ratios $g=V^{IV}_{0}(\beta_2)/V^{IV}_{0}(\beta_2=0.0)$  and 
  $f=V_0^{IS}(\beta_2)/V_0^{IV}(\beta_2)$  at different $\beta_2$ value.  
The vertical 
black line represents the deformation at the lowest energy minimum. See the text for details.}
\end{figure}
Firstly, we use the volume pairing interactions for all pairing channels and assumed an equal pairing strength for the IV channel,  namely,  $V^{pp}_0=V^{nn}_0=V^{IV}_0$ and $x_j=0$ in Eq.(\ref{interaction}). 

As shown in Table.\ref{tab1},  the pp  and nn  pairing gaps are similar values except for $^{64}$Ge 
and $^{68}$Se, so that the pnBCS calculations are performed at each deformation optimizing  the two strengths $V^{IV}_0$ and $V^{IS}_0$  to minimize the root mean squared error between the empirical pairing gaps and the calculated mean pairing gaps.

Fig. \ref{oldratio} (a) shows the ground state energies as a function of the quadrupole deformation 
parameter $\beta_2$ for $^{44}$Ti , calculated by HFB and pnBCS models. The  total energy at the minimum point 
is reduced by about 2 MeV due to the  pn  pairing correlations.
The ratio between the IS and IV pairing strengths,
\be
f=\frac{V^{IS}_0}{V^{IV}_0},
\ee
 is shown in Fig.\ref{oldratio} (b) as a function of
 deformation parameter $\beta_2$. Another ratio  between $V^{IV}_0(\beta)$  at finite deformation and the one at the spherical shape,
 \be
  g=\frac{V^{IV}_0(\beta)}{ V^{IV}_0(\beta=0)},
  \ee
  is also shown in the figure. It can be observed that the ratio $g$ is relatively flat, indicating that the IV pairing strength is changed only slightly in the presence of deformation, namely, the IV pairing is correlated weakly with the deformaiton. 
In contrast, for the ratio $f$, there is a peak at  the oblate deformation $\beta_2\sim -0.2$, while for prolate $\beta_2$ deformation, the ratio $f$ decreases with the increasing of 
the deformation. That is,  the IS pairing correlation is strongly affected by  the  deformation in both  oblate and prolate regions. This characteristic feature is found in all nuclei studied in this article.

\begin{figure}[t]
  \centering
  \includegraphics[scale=0.64,clip=true]{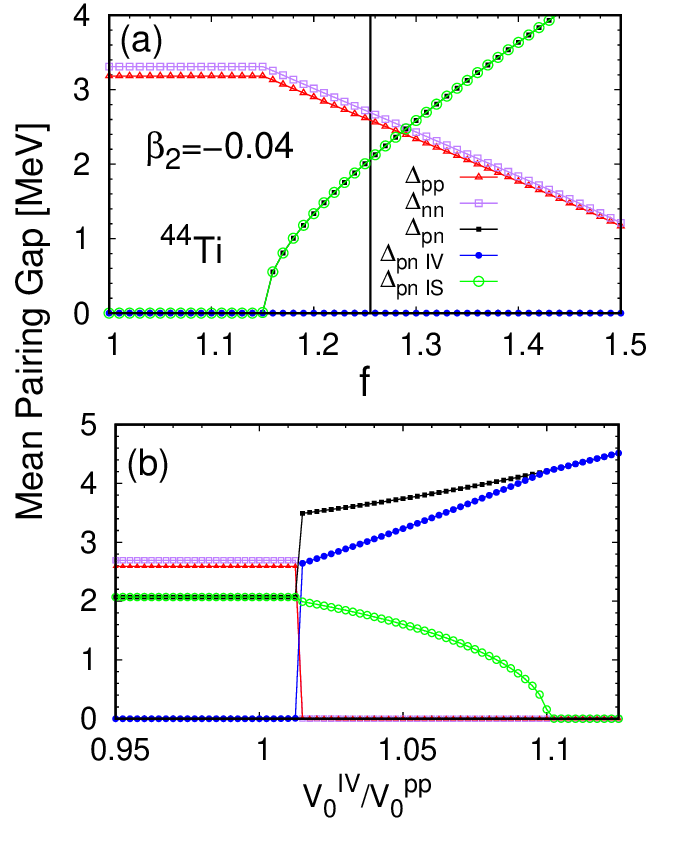}
  \caption{\label{oldgap}\vspace{2.5pt}(a):  {Mean pairing gaps estimated by Eq.(\ref{mgap}) at the energy minimum} as a function of the  ratio $f$ for $^{44}$Ti. The IV pairing strength $V_0^{IV}(V_0^{pp}=V_0^{nn}=V_0^{IV})$ is fixed at the minimum shown in Fig. \ref{oldratio}, whereas the IS pairing strength $V_0^{IS}$ is varied. The  vertical black line represents the position of $f$ that reproduces the empirical energy gap. (b): Mean pairing gaps as a function of the  ratio $V_0^{IV}/V_0^{pp}$ for $^{44}$Ti. The pairing strength $V_0^{pp}(V_0^{pp}=V_0^{nn})$ and $V_0^{IS}$ are set to the same values at the position represented by the solid vertical black line in Fig. (a), whereas the IV pn pairing strength $V_0^{IV}$ is varied. See the text for details.}
\end{figure}

We show the pp, nn, IS and IV pn, and  total pn pairing gaps of $^{44}$Ti in
Fig. \ref{oldgap} obtained at $\beta_2=-0.03$ corresponding to the lowest energy in Fig. \ref{oldratio}(a). In Fig. \ref{oldgap}(a), we fix the strengths of the nn, pp, and IV pn pairing,  and vary only  the IS pairing strength by the ratio $f$. 
As shown in the figure, when $f$ is smaller than  1.15,  there are only pp and nn pairs. While $f$ is larger than 1.15, the IS pn becomes finite and keeps increasing for a larger $f$ value, while the pp and nn pairing gaps
decrease constantly. As a result, the coexistence region of pp, nn, and pn pairing appears and the pp, nn, and pn pairing gaps can be fitted simultaneously. However, the IV pn pairing is absent in the entire region.
In order to check the reason for the absence of IV pn pairing, we fix the strengths of the pp,nn, and IS pn pairing interactions, and  vary only the IV pn pairing strength. The mean pairing gaps are shown in Fig.\ref{oldgap}(b) as a function of the ratio $V_0^{IV}/V_0^{pp}$. As shown in the figure,
when this ratio is smaller than about 1.01, there are only pp, nn, and IS pn pairing and no IV pn pairing, but when the ratio is increased slightly beyond 1.01, the pp and nn pairing suddenly disappeared, and only the IV pn pairing gap appears in the IV pairing correlations. Remarkably, despite this suddenly interchange from  nn and pp to IV pn pairing gaps, the net IV energy gap  exhibits a smooth  dependence on the ratio parameter before and after this transition coexisting with the IV pn pairs. 

This kind of sudden change of the pairing gap is also observed in Ref. \cite{Simkovic} in pnBCS calculations with simple seniority pairing interactions. However, the pp(nn) pairs 
and pn pairs in Ref. \cite{Simkovic} are incompatible. In contrast, the results presented in Fig.\ref{oldgap} (a) show that the IV and IS superfluidities of pp(nn) and pn pairs can coexist. 
The problem of non-coexistence is due to similarities between the proton and neutron wave functions involved in the gap equation,  and also the type of pairing interaction. 
In order to cure this non-coexistence problem, promising models are to use interacting shell model wave functions,  or projected BCS model \cite{Nicu2022}, or to use realistic pairing interactions
for different channels to break the symmetry  in the IV channels 
such as G-matrices \cite{Cheoun2018(2)}.  
In the pnBCS model, 
one can obtain a mean-field solution that allows for the coexistence for the three IV pairs by using different
types of pairing interaction in pp(nn) and IV pn pairing channels.

\subsection{Results with different types of pairing  interactions}
\begin{figure}[t]
  \includegraphics[scale=0.40, clip=true]{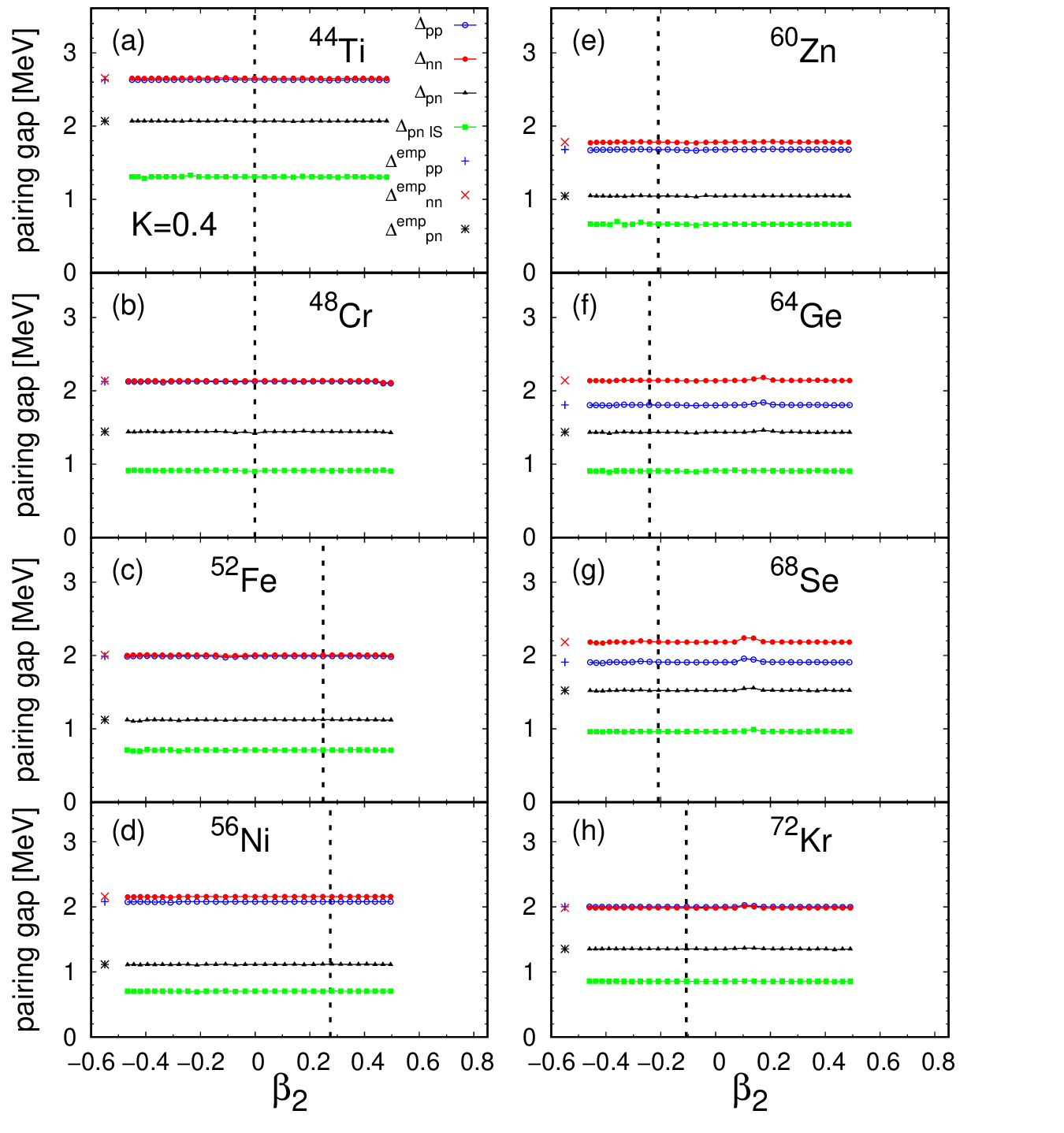}
\caption{\label{DET4}Empirical and mean pairing gaps for $N=Z$ nuclei from $^{44}$Ti to $^{72}$Kr optimized by varying $V_{0}^{pp}$, $V_{0}^{nn}$, $V_{0}^{IV}$ and $V_{0}^{IS}$ as a function of the deformation parameter $\beta_2$ at $K=0.4$. The black dashed line represents the deformation
at the lowest point of energy.}
\end{figure} 
In the previous subsection, we delved into the implications of the volume pairing interaction for both the IV and IS channels, which precludes the simultaneous existence of pp, nn, and pn IV pairing gaps.  
But there is empirical evidence for the exist of IV pn pairing from the energy difference observed between IS and IV states with the same spin-parity \cite{Macchiavelli}.
It might be necessary to modify the current pnBCS model by breaking the isospin symmetry in the isovector (IV) channel, so as to permit the coexistence of pp, nn , and IV pn pairing gaps.

In this subsection we use a surface pairing interaction in the pp and nn channels and a volume pairing interaction in the IS and IV pn pairing channels, which means $x_{pp,nn}=1$ and $x_{IV,IS}=0$ in Eq.(\ref{interaction}). 
In the calculations we do not  constrain any pairing strength  to be the same, but  vary all 
the  parameters $V_{0}^{pp}$, $V_{0}^{nn}$ , $V_{0}^{IV}$ and $V_{0}^{IS}$ to reproduce the three empirical pairing gaps. Because there are four free strength parameters, and  three empirical pairing gaps to be reproduced, one degree of freedom among the four parameters can not be fixed by the experimental data. We then introduce a content parameter of IS pairing gap in  the total pn gap defined as, 
\begin{equation}
\begin{split}
K=(\frac{\overline{\Delta}_{pn}^{IS}}{\overline{\Delta}_{pn}})^{2}.
\end{split}
\end{equation}
to fix the role of a redundant parameter of the $pn$ pairing channel.  The definition of the ratio is defined to be consistent with that of the pn pairing gap,
\be
\Delta _{pn}=\sqrt{(\Delta _{pn}^{IS})^2+(\Delta _{pn}^{IV})^2}
\ee
Fig. \ref{DET4} illustrates the empirical and calculated  pairing gaps for N = Z nuclei from $^{44}$Ti to $^{72}$Kr at $K=0.4$, 
where approximately 40\% and 60\% of the total pn pairing gaps come from the IS and IV channels, respectively. 
Actually, adjusting the four strength parameters $V_{0}^{pp}$, $V_{0}^{nn}$ , $V_{0}^{IV}$ and $V_{0}^{IS}$, all empirical pairing gaps can be reproduced at all the deformations at the fixed ratio $K$ as shown in Fig. 3  for $K=0.4$ case. 
 That is, the coexistence of the four kinds of pairs  can be obtained at any ratio of IS and IV pn pairing gaps,  when different types of pp (nn),  and pn pairing interactions are applied in the calculations.
\begin{figure}[t]
\includegraphics[scale=0.40,clip=true]{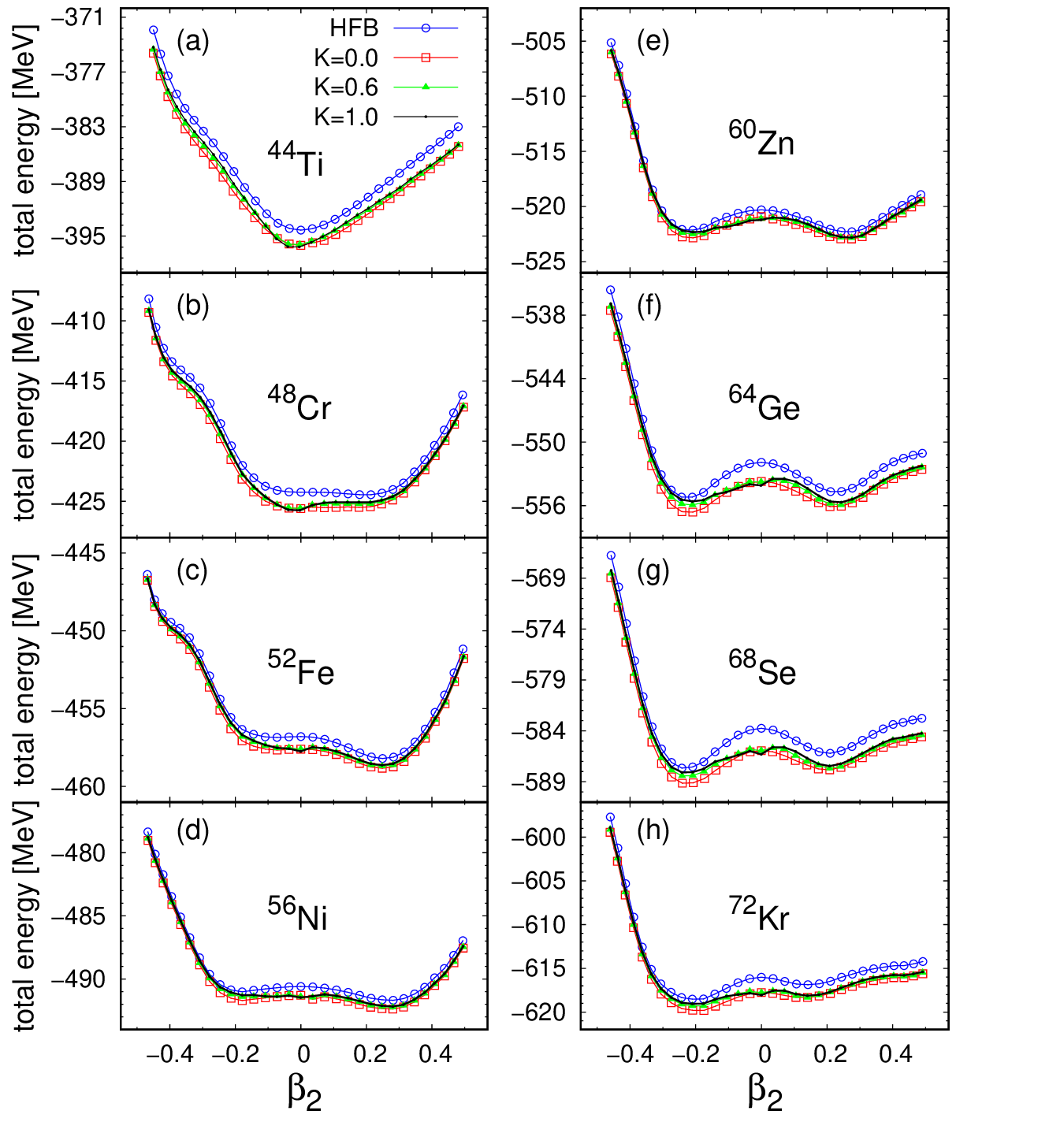}
  \caption{\label{energy}The total energies of ground states as a function of the quadrupole deformation parameter $\beta_2$ for N = Z nuclei from $^{44}$Ti to $^{72}$Kr calculated by the HFB with nn and pp pairings and pnBCS with full pairing interactions at the ratios $K=$0.0,0.6, and 1.0. See the text for details.}
\end{figure}

Fig. \ref{energy} shows the ground state energies with the content ratio of IS pairing set at 0.0, 0.6, and 1.0, as a function of the quadrupole deformation parameter $\beta_2$ for $N=Z$ nuclei, calculated by HFB with pp and nn pairings, and by pnBCS model with all pairing interactions. The lowest total energies are reduced by about 1-3 MeV due to the including of pn  pairing correlations. The total energies for $^{44}$Ti, $^{48}$Cr, $^{52}$Fe, $^{56}$Ni, and $^{60}$Zn are insensitive to the ratio $K$ of IV to IS pn pairing gaps.  That is, the IS and IV superfluidity coexist in the ground states of these nuclei. 
On the other hand,  in the heavier $N=Z$ nuclei, $^{64}$Ge, $^{68}$Se, and $^{72}$Kr, the ground states show IV pn pairing dominance, i.e.,  K=0.0 case corresponds to the lower energies. The estimated deformations at the lowest energies in Fig. \ref{energy} are consistent with the results including pn pairing in Ref. \cite{Cheoun2018(2)}, except for $^{44}$Ti and $^{64}$Ge. Our results demonstrate also that the pn pairing does not have any  significant effect on the deformation of the ground state. 

Table \ref{tab2} presents the four pairing strength values $V_0^{pp}$,$V_0^{nn}$,$V_0^{IV}$ and $V_0^{IS}$ for the deformations correspond to the lowest energy with the ratio$K=0.6$. It is evident that the pairing strengths between pp and nn pairs are always  comparable with less than $10\%$ difference. In contrast, 20$\sim$30 \% larger strength differences between IV and IS pn interactions are induced  by the deformation for nuclei $A \ge 52$.

\begin{figure}[t]
\includegraphics[scale=0.4,clip=true]{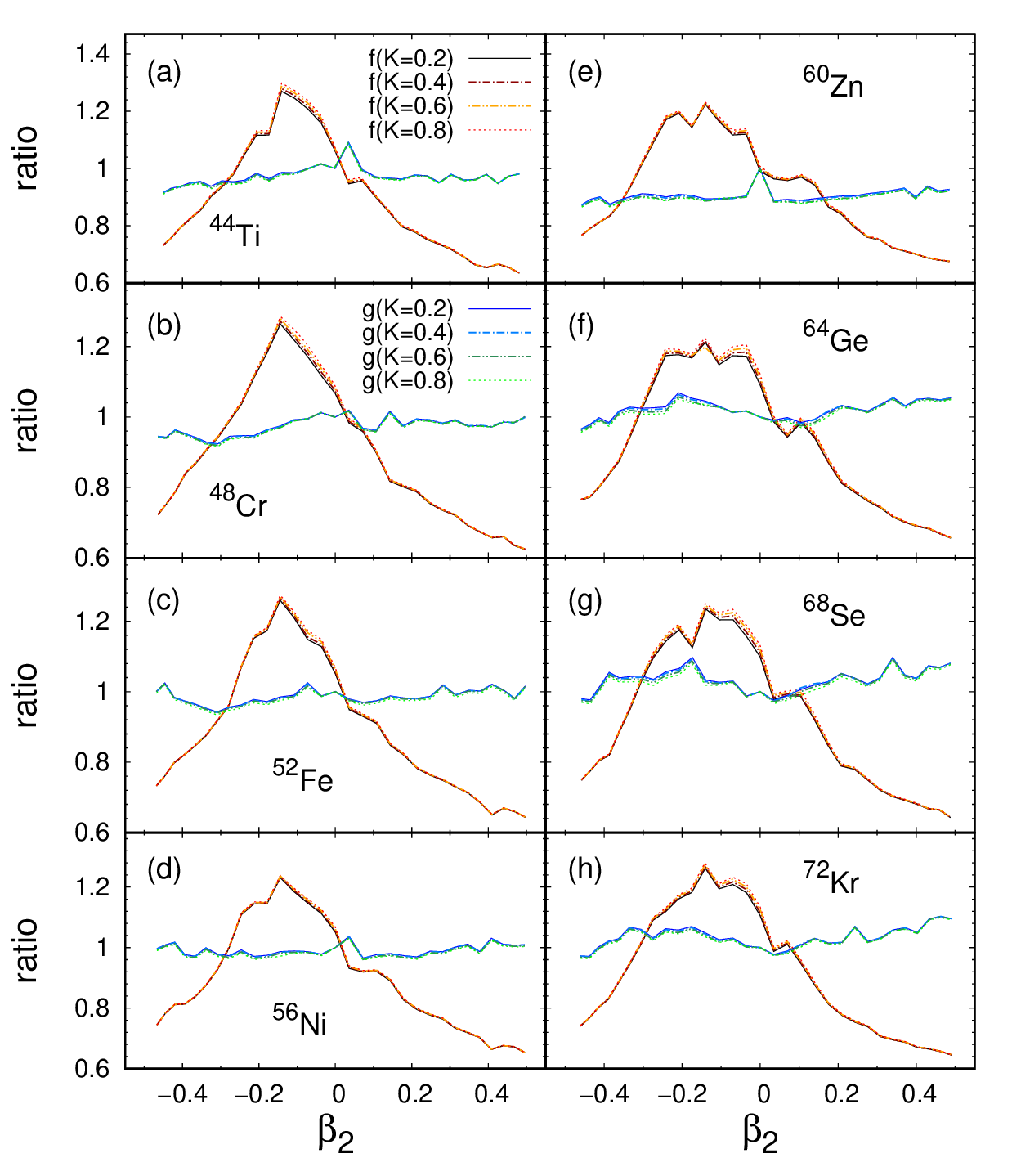}
  \caption{\label{fig:ratio}The same as Fig. \ref{oldratio}(b), but for N = Z nuclei from $^{44}$Ti to $^{72}$Kr at $K=0.2,0.4,0.6,0.8$. At each deformation, the pairing strengths are adjusted to reproduce all experimental pairing gaps.}
\end{figure}

\begin{table}[t]
\caption{\label{tab2}Strength parameters of different types of pairing interactions for N = Z nuclei from $^{44}$Ti to $^{72}$Kr at the ratio K=0.6.  
The volume-type pairing interction is adopted for the pp and nn pairings, while the surface-type one is adopted for the pn pairing.} 
\begin{ruledtabular}
\begin{tabular}{cccccc}
\textrm{Nucleus}&
\textrm{$\beta_2$}&
\textrm{$V^{pp}_0$}&
\textrm{$V^{nn}_0$}&
\textrm{$V^{IV}_0$}&
\textrm{$V^{IS}_0$}\\
\colrule  
$^{44}$Ti&0.00&-693.58&-713.16&-321.79&-346.44\\
$^{48}$Cr&0.00&-658.46&-677.65&-303.87&-329.39\\
$^{52}$Fe&0.25&-677.57&-700.13&-293.66&-224.51\\
$^{56}$Ni&0.280&-691.97&-723.00&-301.67&-231.24\\
$^{60}$Zn&0.27&-673.91&-711.54&-288.17&-219.82\\
$^{64}$Ge&-0.21&-695.63&-761.82&-309.35&-367.22\\
$^{68}$Se&-0.24&-725.48&-785.34&-307.67&-355.72\\
$^{72}$Kr&-0.18&-739.99&-767.92&-311.63&-371.14\\
\end{tabular}
\end{ruledtabular}
\end{table}

From the former subsection we found that the correlation between the IS pairing strength and the deformation.  
We confirm this correlation for different ratios $K$ in the following.
Similar to Fig. \ref{oldratio}(b), the IV and IS pairing strength ratios $f$ and $g$ at $K = 0.2, 0.4, 0.6, 0.8$ are shown in Fig. \ref {fig:ratio}.
It can be seen that for all nuclei, the same correlation between the IS pairing strength and the
deformation exists regardless to the content of IS pn pairing, i.e., the $f$ and $g$ values are almost degenerate for 4 different $K$ values. 
The curves show that the ratio $g$ is relatively flat, whereas the ratio $f$ exhibits a peak at $\beta_2 =-0.1$ for the light $N=Z$ nuclei $^{44}$Ti $\sim ^{60}$Zr, while the peaks show variations in the range of  
$-0.2<\beta_2<0.0$ for heavier $N=Z$ nuclei $^{64}$Ge $\sim ^{72}$Kr. The value of $f$ has maximum value about 1.2 for $\beta_2 \sim -0.2$, whereas  $f$ is reduced to $0.75-0.85$ for prolate $\beta_2 \sim 0.2$.
This implies that for oblate deformation, a stronger pairing strength is required to obtain the empirical pn pairing gap, while for prolate deformation a comparatively weaker pairing strength is enough.

In order to comprehend the correlation between deformation and IS spin-triplet pairing, 
 we study the contribution of each orbit to the pn pairing matrix element in Eq.(\ref{dpn}),
\begin{equation}
\begin{split}
&Im(\Delta_{\alpha p\overline{n}})=\sum \limits_{\beta>0} \int\int Im(u_{\beta 1p}v_{\beta 1n}^*+u_{\beta 2p}v_{\beta 2n}^*)V_{pn(T=0)}  \\&
\times  (\phi_{\alpha\frac{1}{2}p} (r_\perp,z)\phi_{\alpha\frac{1}{2}n} (r_\perp,z)-\phi_{\alpha-\frac{1}{2}p} (r_\perp,z)\phi_{\alpha-\frac{1}{2}n} (r_\perp,z))\\
&\times(\phi_{\beta\frac{1}{2}p} (r_\perp,z)\phi_{\beta\frac{1}{2}n} (r_\perp,z)-\phi_{\beta-\frac{1}{2}p} (r_\perp,z)\phi_{\beta-\frac{1}{2}n} (r_\perp,z))\\
&\times 2\pi  r_\perp dr_\perp dz
\end{split}.
\end{equation}
It can be seen that the IS pairing contribution to the matrix element is related to the difference between the spin-up and spin-down components of the single-particle wave function. 
When the single-particle state is dominated by the spin up or down components, this state has a larger contribution. 
Therefore, we define the quantity $P_{\alpha}$ to estimate the contribution of a single particle state $\alpha$ to the pairing matrix element:
\begin{equation}
\begin{split}
P_{\alpha}=\left \vert 2 \pi\int (\phi_{\alpha\frac{1}{2} p}\phi_{ \alpha \frac{1}{2} n}-\phi_{\alpha-\frac{1}{2} p}\phi_{\alpha-\frac{1}{2} n}) r_\perp dr_\perp dz\right \vert
\end{split}
\end{equation}
If a single-particle state $\alpha$ is dominated by spin-up (or -down) component, $P_{\alpha}$ approaches $1$, it has a substantial contribution to IS pairing 
matrix element. On contrast, when the state $\alpha$ has  similar spin up and down components, $P_{\alpha}$ approaches $0$, and consequently gives small contribution to IS pairing matrix element.

To elucidate the shape of individual single-particle state evolving with the deformation of nuclei, we introduce the deformation index $\beta^{\alpha}_{2}$ for each single-particle state, defined as:
\begin{equation}
\begin{split}
\beta^{\alpha}_{2}=\sqrt{\frac{\pi}{5}}\frac{\langle \Phi_{\alpha,n} \vert 2z^2-r_{\perp}^2\vert \Phi_{\alpha,n}\rangle  }{\langle \Phi_{\alpha,n} \vert z^2+r_{\perp}^2\vert \Phi_{\alpha,n}\rangle},
\end{split}
\end{equation}

\begin{figure}[t]
\includegraphics[scale=0.65,clip=true]{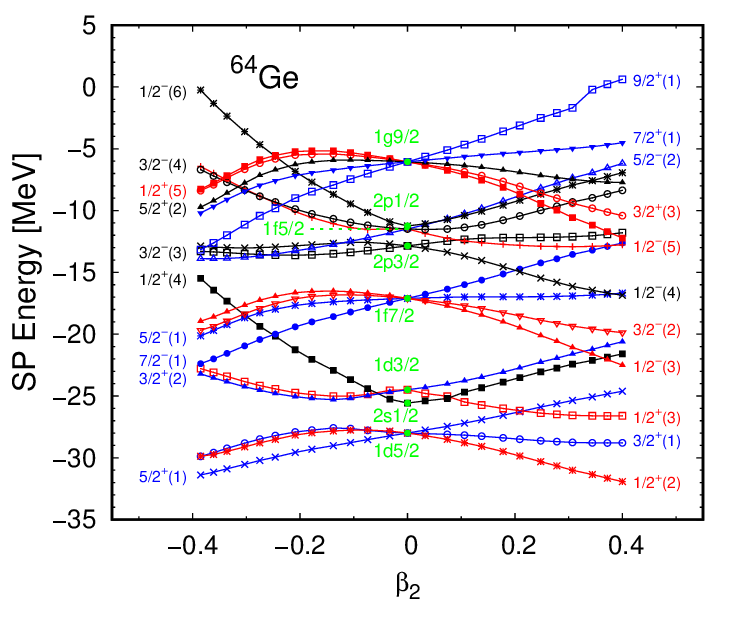}
  \caption{\label{SPE} The neutron single particle energy of $^{64}$Ge as a function of deformation parameter $\beta_2$. Single-particle states are shown in blue, black, and red colors correspond to single particle states with larger, medium, and smaller $\Omega$ values coming from  the same energy levels at the spherical limit.}
\end{figure}

Taking $^{64}$Ge as an example, we focus on 21 single particle states within the 1-2 major shells below and above the Fermi surface, which make a major contribution to the gap eqautions. 
Figure \ref{SPE} presents a Nilsson-type diagram of the 21 states, depicting the deformation-dependent evolution of individual energy levels.
Since proton and neutron wave functions exhibit almost identical behaviors in the single-particle diagram  in $N=Z$ nucleus, we present only the single-particle energies of neutrons.

\begin{figure}
  \includegraphics[scale=0.36,clip=true]{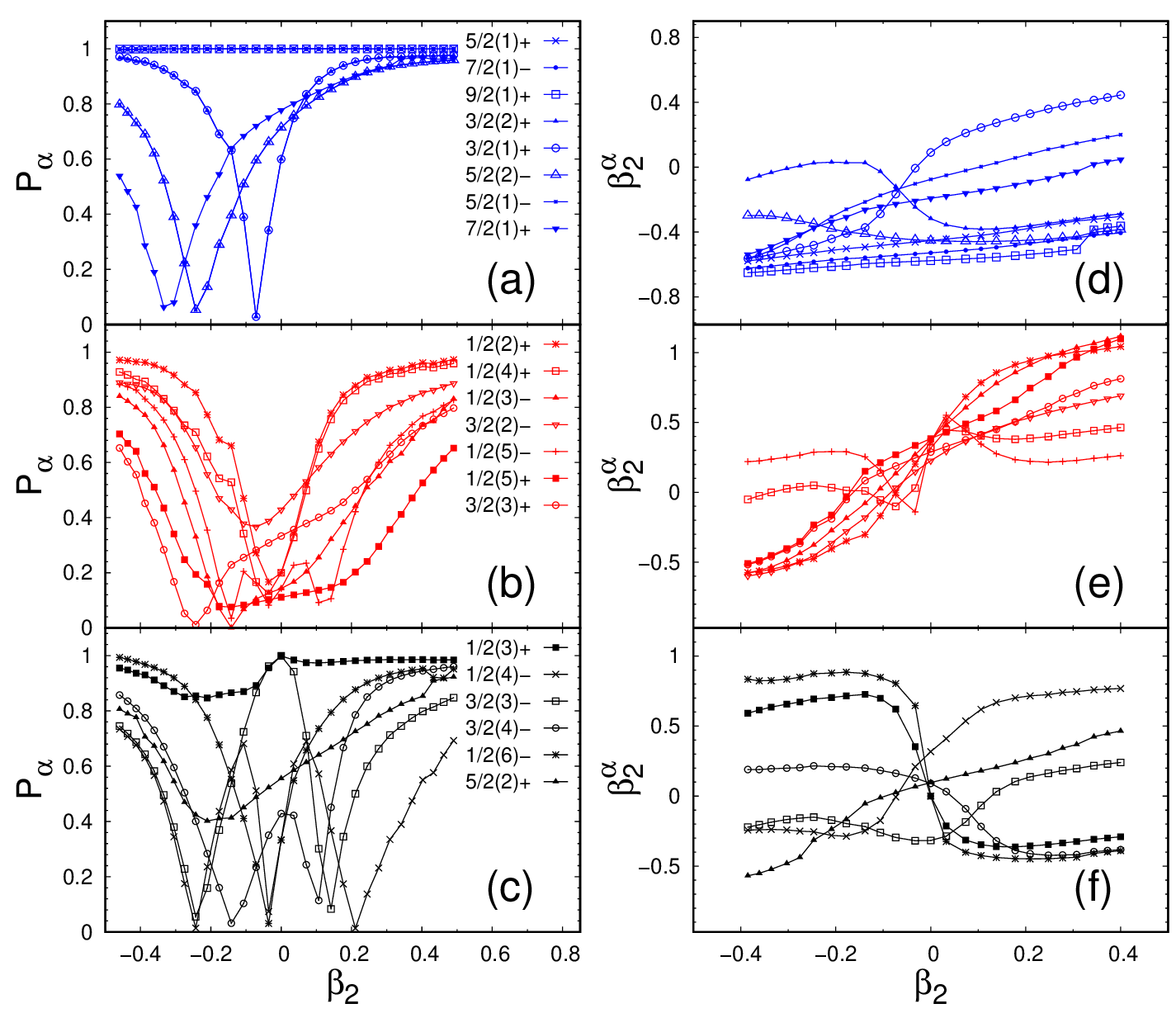}
  \caption{\label{fig:FSL}. 
  $P_{\alpha}$ values and the single-particle deformation index $\beta^{\alpha}_2$ in $^{64}$Ge as a function of the nuclear deformation $\beta_2$.
 The $P_{\alpha}$ values are shown in panels (a), (b), and (c), while the $\beta^{\alpha}_2$ values are shown in panels (d), (e), and (f), respectively. The labels and colors of single particle states are the same as in Fig. \ref{SPE}. See the text for details.}
\end{figure}

$P_{\alpha}$ and $\beta_2^\alpha$ of the 21 states as a function of the deformation parameters $\beta_2$ are illustrated in Fig.\ref{fig:FSL}. 
Among the 21 states  in Figs. \ref{fig:FSL} (a), (b), and (c), 17 states show very small  $P_{\alpha}$ values  
 in the region $-0.3 < \beta_2 < 0.0 $,  
As a result, a stronger IS pairing strength is required to fit the empirical pairing gap in the oblate deformation region. 
For prolate deformation, as the $P_\alpha$ values of most states are going up with the increasing of the deformation, smaller IS pairing strengths will be enough to produce the pn pairing gap.
For larger prolate or oblate deformation with $|\beta_2|>0.5$,  
the deformed states are well represented by the Nilsson quantum numbers, where the spin-$z$ component becomes a good quantum number. In this case, as a natural consequence, a much smaller $f$ is enough to reproduce the empirical pn pairing gap as illustrated in Figs. 1(b) and Fig.\ref{fig:ratio}.

\begin{figure} [t] \centering
\includegraphics[width=3.5in,scale=0.3,clip=true]{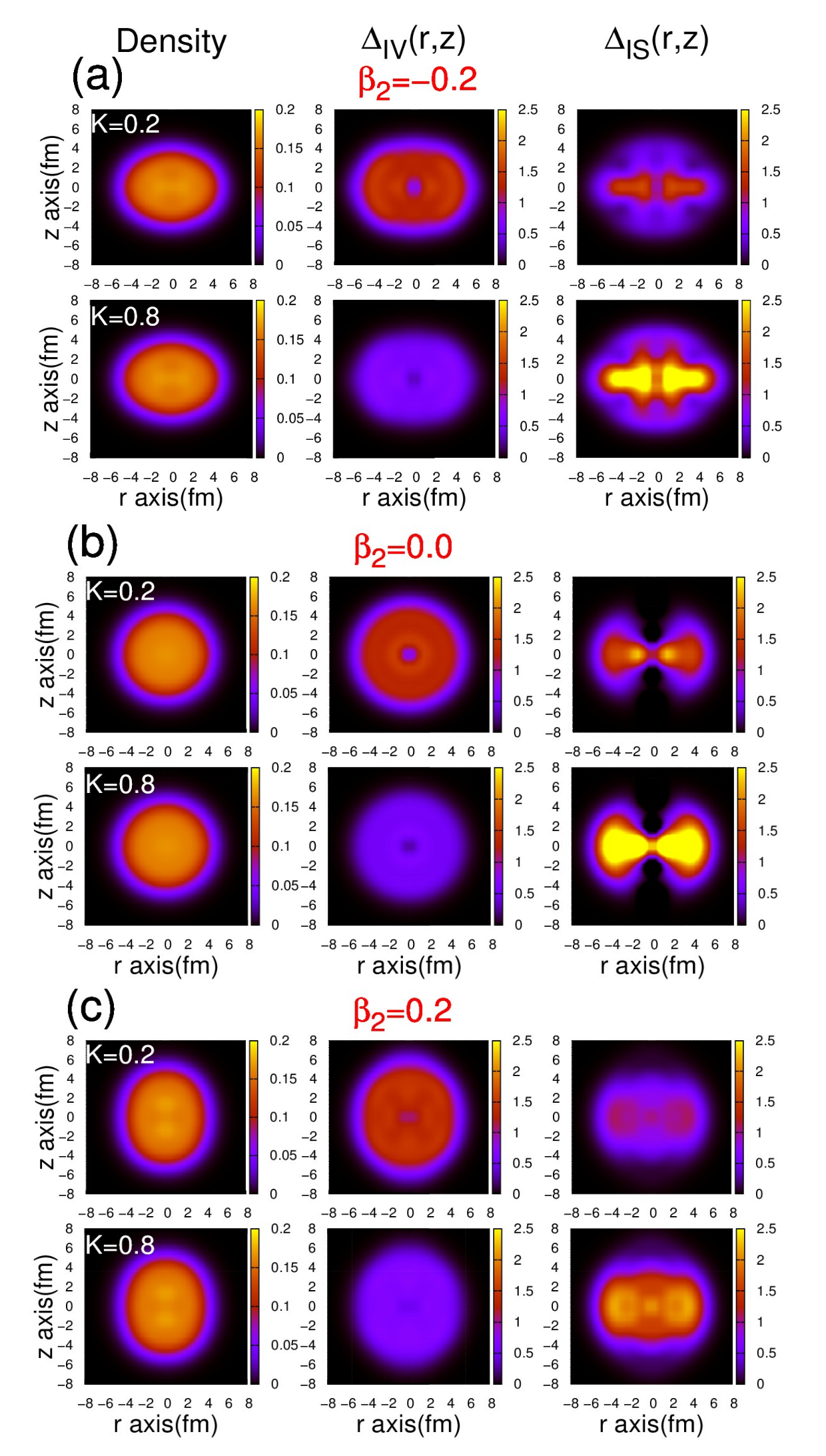}
  \caption{\label{fig:field}The normal density (left) and the abnormal density distributions of IV (middle) and IS (right) pn pairings  in $^{64}$Ge for the IS and IV mixing ratio $K=0.2$ and $0.8$, respectively.  The deformation parameter is taken to be  $\beta_2=-0.2$, $0.0$, and $0.2$, 
   for  panels (a), (b), and (c), respectively.}
\end{figure}
Fig. \ref{fig:field} shows the normal density,  and the abnormal density distributions of IV and IS pn pairing fields of $^{64}$Ge.  
It is evident that the  shape of abnormal IV pn pairing density is similar to that of normal density since all wave functions contribute to the abnormal IV spin-singlet density in the same way as to
the normal density except the $u$ and $v$ factors.  
In contrast, the IS pn pairing field exhibits oblate shape regardless of the nuclear deformation. 
We have checked the IS pn pairing fields of the other seven nuclei, and found that they all exhibit similar oblate shape regardless to the deformation of normal density.
  
To understand this unexpected oblate dominance in the IS abnormal density,  we study the deformation dependence of the index $\beta_2^\alpha$ in Eq. (15).
As illustrated in Figs.\ref{fig:FSL} (d), (e), and (f), in the case of oblate  deformations,  
the $\beta^\alpha_2$  of 15 states are negative among 21 states. As a natural consequence,  the dominant contributions to the  isoscalar pairing  originate from the states with negative shape index, and the abnormal IS density becomes oblate. 
In the case of prolate deformation, 
8 states retain  still negative shape index $\beta_2^\alpha$ despite to the fact that the nucleus exhibits a prolate deformation as a whole. Furthermore, the $P_{\alpha}$ for the  states with positive shape index $\beta_2^\alpha$  remain smaller than those with negative  values.
As a net, the  states with negative index $\beta_2^\alpha$ contribute more that those with positive index and make the IS abnormal density to be weakly oblate.  
The same feature is also observed in the spherical case.

\section{summary}\label{section:IV}
We studied the competition between IS and IV pairing condensation in eight $N=Z$ even-even  nuclei from $^{44}$Ti to $^{72}$Kr 
 by using an axially symmetric deformed HF+
pnBCS model.  We adopted   a  
 Skyrme EDF SGII with the contact  pairing interactions whose strengths adjusted to reproduce the empirical pairing gaps.
The coexistence ratio of IV to IS pairing condensations is changed by the pn pairing strengths to check possible co-existence and also the 
dominant phase of superfluidity.  
Among the eight N=Z nuclei, we found that the ground states of $^{64}$Ge, $^{68}$Se, and $^{72}$Kr show the dominance of IV  superfluidity. On the other hand, 
 other five nuclei $^{44}$Ti, $^{48}$Cr, $^{52}$Fe, $^{56}$Ni and $^{60}$Zn  exhibit the coexistence of IV spin-singlet and IS spin-triplet pairing condensations.
 We do not find any nucleus which shows dominant IS condensation phase in this mass region $A=44\sim 72$. 
 
We pointed out the correlation between 
the IS pn pairing strength and the nuclear deformation;  for oblate deformation with $-0.3<\beta_2<0.0$, a stronger IS pn pairing strength is needed to reproduce the empirical pairing gaps. In contrast, for prolate deformation,  a weaker IS pn pairing strength is enough to obtain the empirical gap.  This feature can be  understood looking at the  spin structure of the deformed single-particle states, namely,   
for oblate deformation with $-0.3<\beta_2<0.0$, the spin-up and spin-down components of most of the active single-particle states 
are mixed up with competitive amplitudes for the IS gap equation and cancel each other largely in the IS pairing matrix element.  In contrast, 
for prolate deformation, the active states  are  dominated by either spin up or spin down components and no cancelation occurs in the gap equation. We found also that the contributions to IS abnormal density are dominated by the oblate-type single-particle states regardless to the deformation of ground state density, i.e., the IS abnormal density shows always oblate deformation no matte how the deformation of normal density is prolate, spherical or prolate.

\end{document}